%% file: main.tex
\documentclass[runningheads,a4paper]{llncs}

\input{flow/flow}

\fxsetup{status=draft}
\fxsetup{theme=color, mode=multiuser} \FXRegisterAuthor{me}{ame}{\color{red}Me}
\newcommand{\orcid}[1]{\href{https://orcid.org/#1}{\includegraphics[scale=0.02]{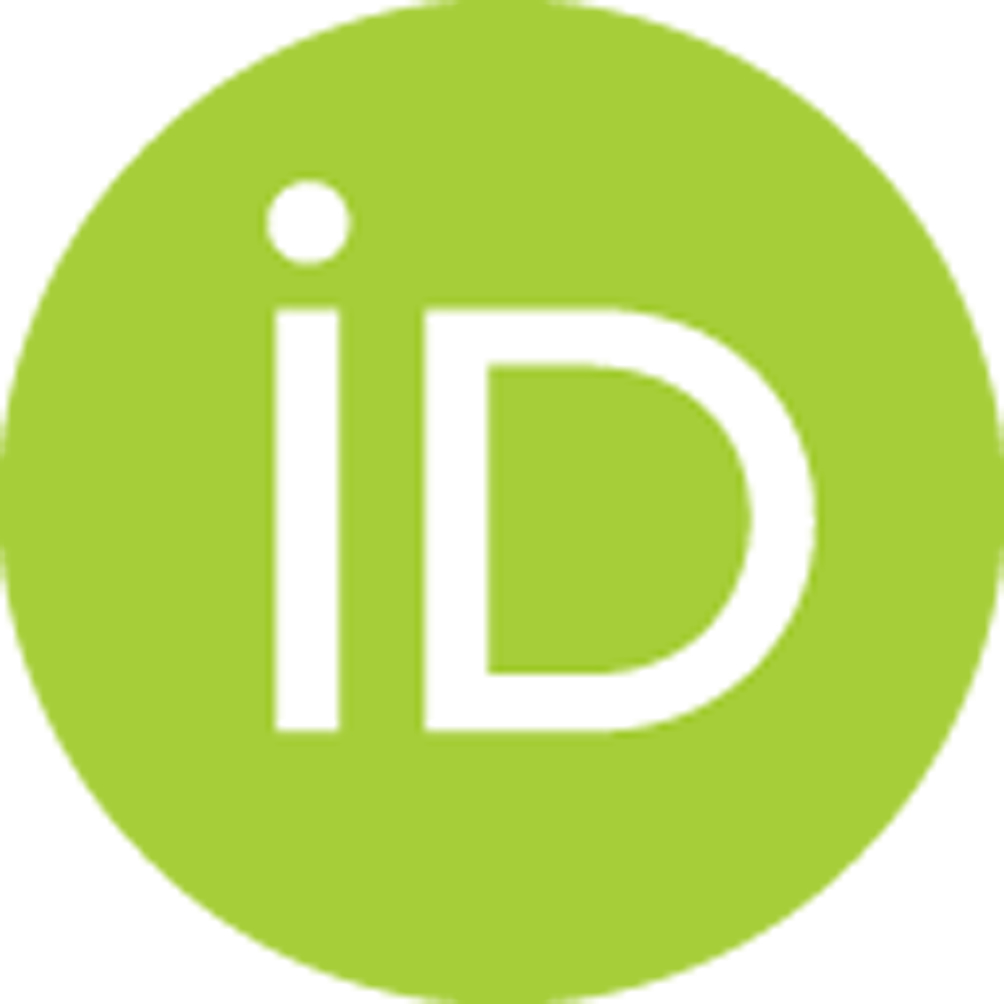}}} 

\hypersetup{
    colorlinks=true
}

\input{head/title}

\input{head/author_list} 

\begin{document}
\mainmatter
\maketitle
\input{head/abstract}

\input{head/keywords}


\input{body/body}





\bibliographystyle{plainnat}
\bibliography{bibliography.bib}
\newpage
\appendix
\section*{Appendix}
\input{appendix/appendix}

\end{document}

%% file: flow/flow.tex
\usepackage[top=2.5cm, bottom=2.5cm, left=2.5cm, right=2.5cm]{geometry}
\usepackage{authblk} 

\usepackage{amsmath}
\usepackage{amssymb}
\usepackage{graphicx}
\usepackage{booktabs}
\usepackage{listings}
\usepackage{xcolor}
\usepackage{hyperref}
\usepackage{caption}
\usepackage{subcaption}

\usepackage{url}
\usepackage{color}
\usepackage{epsfig}
\usepackage{float}
\usepackage{tabularx}

\usepackage[nomargin,inline,marginclue,draft]{fixme}

\usepackage{import}
\usepackage{currfile}

\usepackage[numbers]{natbib}
\usepackage{hyperref}

\usepackage{varioref}

\usepackage{cleveref}

\usepackage[nolist]{acronym}
\input{flow/acronyms}

\input{flow/commands}

%% file: flow/acronyms.tex
\begin{acronym}
\acro{CNN}{Convolutional Neural Network}
\acro{DL}{Deep Learning}
\acro{DSC}{Dice similarity coefficient}
\acro{GAN}{Generative Adversarial Network}
\acro{ML}{Machine Learning}
\acro{MRI}{Magnetic Resonnance Imaging} 
\acro{PGT}{Peak Ground Truth}
\acro{RWMP}{Real World Model Performance}
\acro{NN}{Neural Network}
\acro{SSIM}{Structural Similarity Index Measure}
\acro{PSNR}{Peak-Signal-To-Noise-Ratio}
\acro{MSE}{Mean-Square-Error} 
\acro{RMSE}{Root Mean-Square-Error} 
\acro{NIH}{National Institutes of Health}
\acro{CSF}{Cerebrospinal Fluid}
\acro{MAE}{Mean Absolute Error} 

\end{acronym}

%% file: flow/commands.tex
\newcommand{\eac}[1]{\emph{\ac{#1}}}

%% file: head/title.tex
\title{The Brain Tumor Segmentation (BraTS) Challenge: \\ \emph{Local Synthesis of Healthy Brain Tissue via Inpainting}}
\titlerunning{BraTS Inpainting Challenge}

%% file: head/author_list.tex
\author[1,2,3,73]{Florian  Kofler}
\author[2]{Felix  Steinbauer}
\author[2,4]{Felix  Meissen}
\author[2,4]{Robert  Graf}
\author[5]{Stefan K Ehrlich}
\author[6,7]{Annika  Reinke}
\author[1,8]{Eva  Oswald}
\author[2,9]{Diana  Waldmannstetter}
\author[2,4]{Florian  Hoelzl}
\author[2,10]{Izabela  Horvath}
% ...
\affil[1]{Helmholtz AI, Germany}
\affil[2]{Technical University of Munich, Germany}
\affil[3]{Helmholtz Munich, Germany}
\affil[4]{Klinikum Rechts der Isar, Munich, Germany}
\affil[5]{SETLabs Research GmbH, Healthcare Technologies, Munich, Germany}
\affil[6]{German Cancer Research Center (DKFZ) Heidelberg, Division of Intelligent Medical Systems, Germany}
\affil[7]{German Cancer Research Center (DKFZ) Heidelberg, HI Helmholtz Imaging}
\affil[8]{Institute of Clinical Neuroimmunology, Faculty of Medicine, Ludwig Maximilian University, Munich, Germany}
\affil[9]{University of Zurich, Switzerland}
\affil[10]{Helmholtz Research Center, Munich, Germany}
% ...
\affil[73]{Biomedical Image Analysis \& Machine Learning, Department of Quantitative Biomedicine, University of Zurich, Switzerland}

%% file: head/abstract.tex
\begin{abstract}
    A myriad of algorithms for the automatic analysis of brain MR images is available to support clinicians in their decision-making.
    For brain tumor patients, the image acquisition time series typically starts with an already pathological scan.
    This poses problems, as many algorithms are designed to analyze healthy brains and provide no guarantee for images featuring lesions.
    Examples include but are not limited to algorithms for brain anatomy parcellation, tissue segmentation, and brain extraction.
    To solve this dilemma, we introduce the BraTS inpainting challenge.
    Here, the participants explore inpainting techniques to synthesize healthy brain scans from lesioned ones.
    The following manuscript contains the task formulation, dataset, and submission procedure.
    Later, it will be updated to summarize the findings of the challenge.
    The challenge is organized as part of the ASNR-BraTS MICCAI challenge.
\end{abstract}

%% file: head/keywords.tex
    \keywords{BraTS, challenge, brain, tumor, segmentation, machine learning, artificial intelligence, AI, infill, in-painting, \emph{challenge specific keyword}}

%% file: body/body.tex

\input{body/introduction/introduction}

\input{body/related/related}
\input{body/definition/definition}

\input{body/2023/2023}

\input{body/2024/2024}

\input{body/discussion/discussion}

\input{body/acknowledgement/acknowledgement}

%% file: body/introduction/introduction.tex
\section{Introduction}
Traditionally, the BraTS challenge \citep{menze2014multimodal,bakas2017advancing,bakas2018identifying,baid2021rsna} was mainly concerned with glioma segmentation.
In 2023, BraTS evolved into a cluster of challenges covering the segmentation of various tumor entities and further towards handling missing data, and tumor image processing tasks.
With this, the BraTS inpainting challenge was introduced.

Unlike the original segmentation challenge, the inpainting challenge focuses on the \emph{local synthesis} of brain \eac{MRI}.
We define \emph{local synthesis} as image synthesis within the brain area affected by a lesion, and the task is to \emph{fill in} i.e., \emph{inpaint} this area with image contrasts that would represent healthy tissue or background\footnote{In this manuscript, we use the term \emph{infilling} synonymously for \emph{inpainting}.}.


Local inpainting enables a synthetic “removal” of the tumor area from the image, which, in turn, offers means for applying standard brain image segmentation algorithms in tumor patients (i.e. brain parcellation algorithms) – without any extra need for modification.
This might bring a deeper understanding of the relationship between different brain tumor regions (brain parcellation) and abnormal brain tissue (brain tumors).
Brain tissue segmentation would be highly important for downstream tasks such as brain tumor modeling \citep{ezhov2021geometry,ezhov2023learn}.
Furthermore, inpainting would allow us to deal with local artifacts, such as those arising from B-field inhomogeneities that occasionally degrade brain tumor images.

The clinical impact of our inpainting challenge will be in enabling the use of BraTS algorithms in case of non-standard imaging protocols \citep{villanueva2017current}, and in directly using brain parcellation tools, such as \citep{jenkinson2012fsl,fischl2012freesurfer}, for treatment planning and the localization of areas of risk.

Complementing the clinical perspective of improved algorithmic deployment, we – as the BraTS organizers – also envision technical advancements from a post-challenge use of the algorithms of this challenge: we will be able to further enrich the BraTS (training) data set by offering different whole brain parcellation masks from established neuroimaging tools for all BraTS cases.
Both contributions will enable new lines of technical and neuroimaging research.


%% file: body/related/related.tex
\section{Related work}
Inpainting is a well-researched problem in the computer vision community.
Numerous algorithms are proposed to realistically infill missing parts for 2D natural images \citep{zhao2021comodgan,romero2022ntire,Guo_2021_ICCV,yu2018free,yu2018generative} and a smaller number for brain images \citep{rouzrokh2022multitask}.

However, whether such algorithms generalize to three-dimensional inpainting of \eac{MRI} scans remains an open research question.
To explore this, we provide a forum to discuss and a platform to benchmark the newly developed methods.

%% file: body/definition/definition.tex
\section{Challenge Definition}
\label{sec:definition}

\import{\currfiledir}{task}


\import{\currfiledir}{data/data}
\import{\currfiledir}{masks}

\import{\currfiledir}{manual_curation}

\import{\currfiledir}{participation}
\import{\currfiledir}{baseline}
\import{\currfiledir}{evaluation/evaluation}

%% file: body/definition/task.tex
In the BraTS inpainting challenge, we encourage participants to offer algorithmic solutions to synthesize 3D healthy brain tissue to replace pathologic brain areas, c.f.
\Cref{fig:challenge_task}.
Given an image and masks corresponding to inpainting targets, the task is to infill the masked pathological areas with healthy tissue.


\input{body/definition/fig/challenge_task}

%% file: body/definition/fig/challenge_task.tex
\begin{figure}[htbp]
        \centering
        \includegraphics[width=1.0\linewidth]{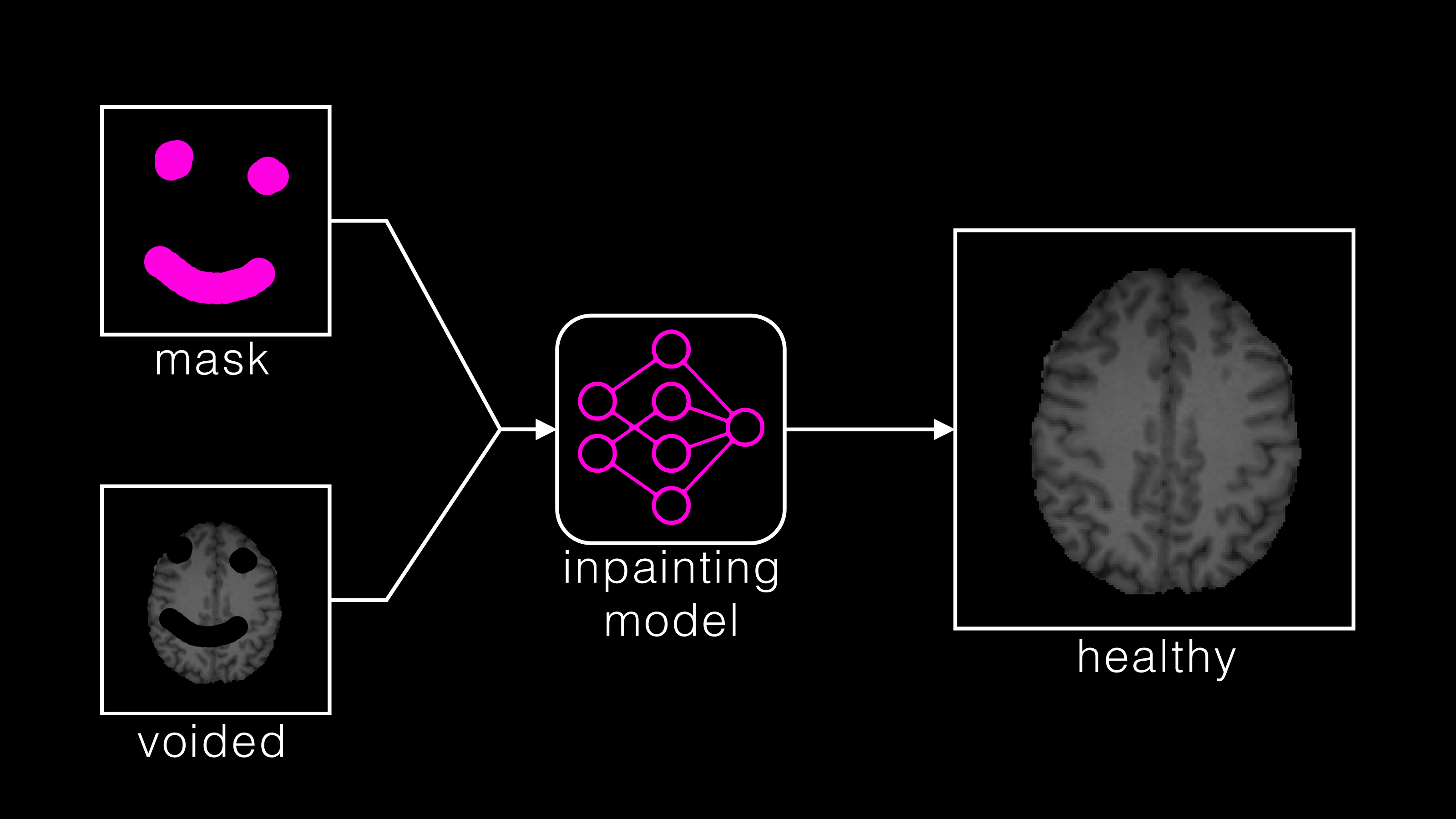}
        \caption{
                \textbf{Challenge Task:}
                Given a set of pathological and incomplete T1 \eac{MRI} images with specified (masked) voided areas, the participants are tasked to synthesize healthy versions of the pathological brains.
                We cast the problem of synthesizing the healthy tissue as inpainting within the tumor area.
        }
        \label{fig:challenge_task}
\end{figure}

%% file: body/definition/data/data.tex
\subsection{Data}\label{sec:data}
The challenge operates on data from the \href{https://www.synapse.org/Synapse:syn51156910/wiki/621282}{BraTS lesion segmentation challenges}.
The tumor segmentation challenges provide retrospective collections of brain tumor \eac{MRI} scans acquired from multiple institutions under standard clinical conditions, but with various equipment and imaging protocols \citep{menze2014multimodal,bakas2017advancing,bakas2018identifying,labella2023asnrmiccai}.
Additionally, multi-class tumor delineations approved by expert neuroradiologists are available.

While the lesion segmentation challenges provide multi-modal image data, the inpainting challenge exclusively employs T1 scans.
The motivation for this is two-fold:
First, to reduce the computational barrier for participants; second, to derive algorithms that easily generalize towards other brain pathologies, as a native T1 scan is obtained in \eac{MRI} protocols for most pathologies.

Following the paradigm of algorithmic evaluation in machine learning, the BraTS inpainting challenge data are divided into training, validation, and testing datasets.
Since no healthy ground truth tissue is available for the tumor regions,
surrogate inpainting masks are generated in the healthy part of the tumor and can be used for training supervised infill algorithms, as well as in the final evaluation.

To maintain a low barrier to participation, we provide participants with suggestions of healthy regions they can use to train their algorithms on (training set).
This generation protocol for healthy masks is described in more detail in Section \Cref{sec:healthy_region_protocol} and is publicly available.
Participants are highly encouraged to iterate on it and use it to extend the quality and quantity of their training dataset.
To enable this process, we provide the original T1 scans for all brains in the training dataset, c.f. \Cref{fig:data_provided}.

For inference, the algorithms are provided with the inpainting masks and the T1 scans with voided masked areas.
We apply this voiding, defined as setting voxels to an intensity of zero, to avoid revealing target information.
The testing data are kept hidden from the participants at all times.
The non-voided T1 images are provided only for training purposes.
A more detailed tutorial on an exemplary training and submission pipeline, including dataset handling and generation, is provided on the \href{https://github.com/BraTS-inpainting}{challenge GitHub page}.

\input{body/definition/data/fig/data_fig}


%% file: body/definition/data/fig/data_fig.tex
\begin{figure}[htbp]
    \centering
    \includegraphics[width=0.8\linewidth]{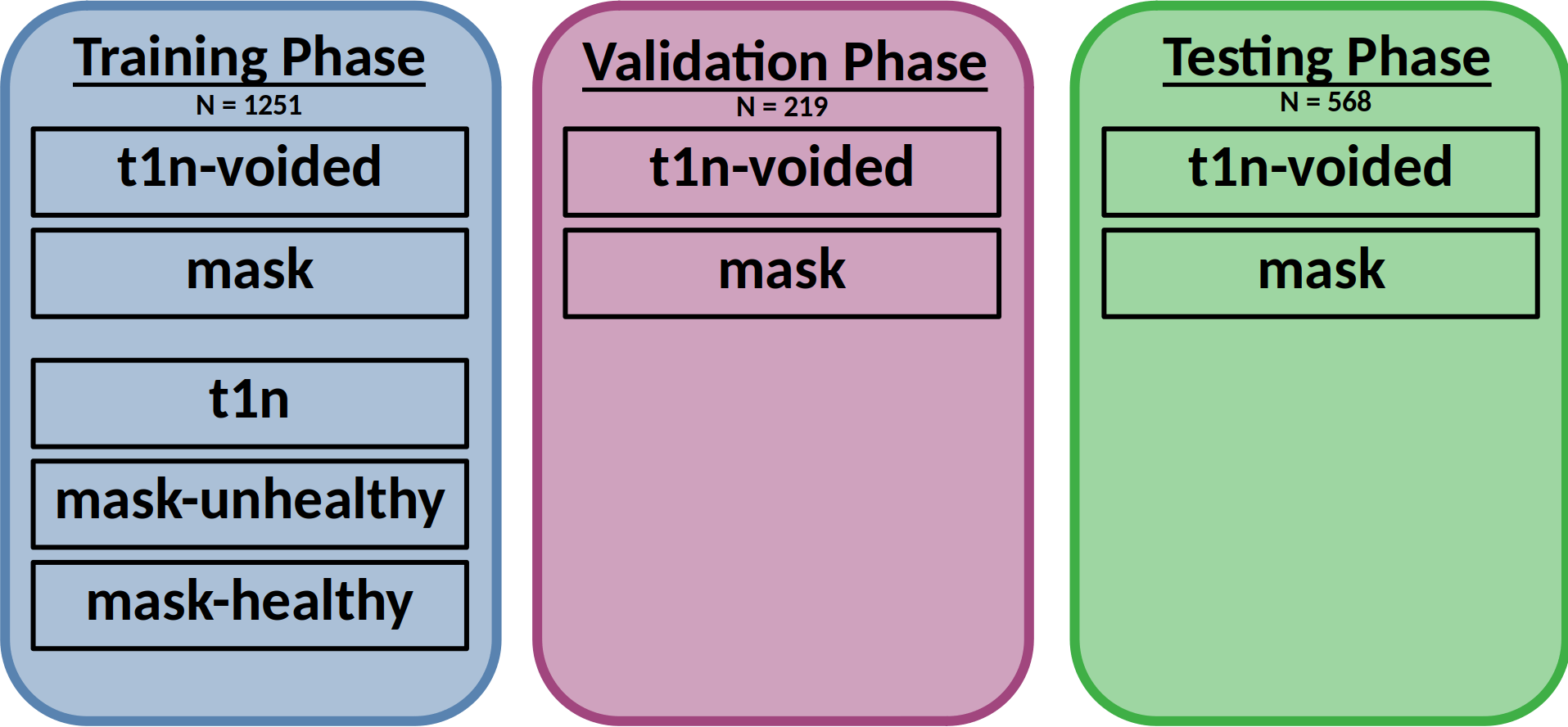}
    \caption{
        \textbf{Dataset figure:} Here, we depict the properties of the available imaging modalities at each stage of the challenge.
        During all phases, we provide a T1 image with voided regions and a corresponding void mask.
        Only, during training we provide the corresponding whole T1 image with an additional mask for the tumor and healthy area.
        These masks are provided to help participants with sampling healthy tissue.
        However, we explicitly encourage participants to experiment with better sampling strategies.
    }
    \label{fig:data_provided}
\end{figure}

%% file: body/definition/masks.tex
\subsection{Healthy Inpainting Mask Generation} \label{sec:healthy_region_protocol}
We work on T1 sequence images and provide two types of masks to participants.
The first type of mask delineates areas which are likely to be affected by the tumor or mass effect.
In contrast, the second type masks areas are likely to be healthy.
Consequently, non-masked areas can be considered as \emph{contested territory}.

To create the masks for healthy brain regions, we sample from existing tumor shapes and place masks in regions distant from the tumor.
This enables training and evaluating the inpainting algorithms on realistic shapes.
It is important to note that, unlike tumors, the inpainting targets can overlap with ventricles, \eac{CSF}, and out-of-brain background, as the participants' algorithms are supposed to correctly inpaint all such areas.
 \Cref{fig:infill_annotations} visualizes the procedure to generate inpainting masks.

 \input{body/definition/fig/infill_annotations}

First, a pool of potential masks is created by taking all disconnected compartments of existing tumor annotations, which consist of at least 800 voxels (\Cref{fig:infill_annotations}, \textbf{a}).
A pool of 1429 masks is extracted from 1251 different brains taken from the official BraTS training dataset.
In the next step, a mask is randomly chosen from the previously created pool biased by the size of the tumor that is already present in the respective brain (\Cref{fig:infill_annotations}, \textbf{b}).
Small masks are chosen for big tumors and vice versa.
This is achieved by selecting masks on the other side of the size distribution of the existing tumor.
For example, if our existing tumor annotation size is in the $80^{th}$ percentile (big tumor), we take a mask around the $20^{th}$ percentile (small mask).
Respectively, if the tumor size is around the median (e.g., $55^{th}$ percentile), a healthy mask with a similar size ($45^{th}$ percentile) is used.

After a mask is chosen from the pool, it is transformed using random mirroring (\Cref{fig:infill_annotations}, \textbf{c}) (mirror probability of 50\% independently in each dimension) and rotation (\Cref{fig:infill_annotations}, \textbf{d}). For rotation, a random angle between 0° and 360° is used for both the \emph{X-Y} and the \emph{Y-Z} planes.
After the transformations, the mask is placed at a semi-random position (\Cref{fig:infill_annotations}, \textbf{e}) based on the distance to the existing tumor; Two random points are sampled within the brain (non-zero T1) but not in the tumor.
According to Euclidean voxel distance, the point furthest away from the tumor is chosen as the position for the healthy tissue mask.

Afterward, the mask's validity is checked based on a minimal distance (\Cref{fig:infill_annotations}, \textbf{f}) to the tumor annotation of 5 voxels euclidean distance and that at most 25\% overlap (\Cref{fig:infill_annotations}, \textbf{g}) with the background (zero T1).
If one of these conditions is violated, the sampling process with subsequent transformation and placement is repeated.
This mask sampling procedure is carried out until one valid configuration is found.
The resulting healthy mask, as well as the (dilated) tumor mask, is provided to participants.

The above-described sampling approach is applied only once per brain for the training, validation, and testing dataset. Of course, more than one mask can be generated per brain.
We encourage participants to improve this procedure and generate more healthy in-painting masks for training purposes.

%% file: body/definition/fig/infill_annotations.tex
\begin{figure}[htbp]
    \centering
    \includegraphics[width=1\linewidth]{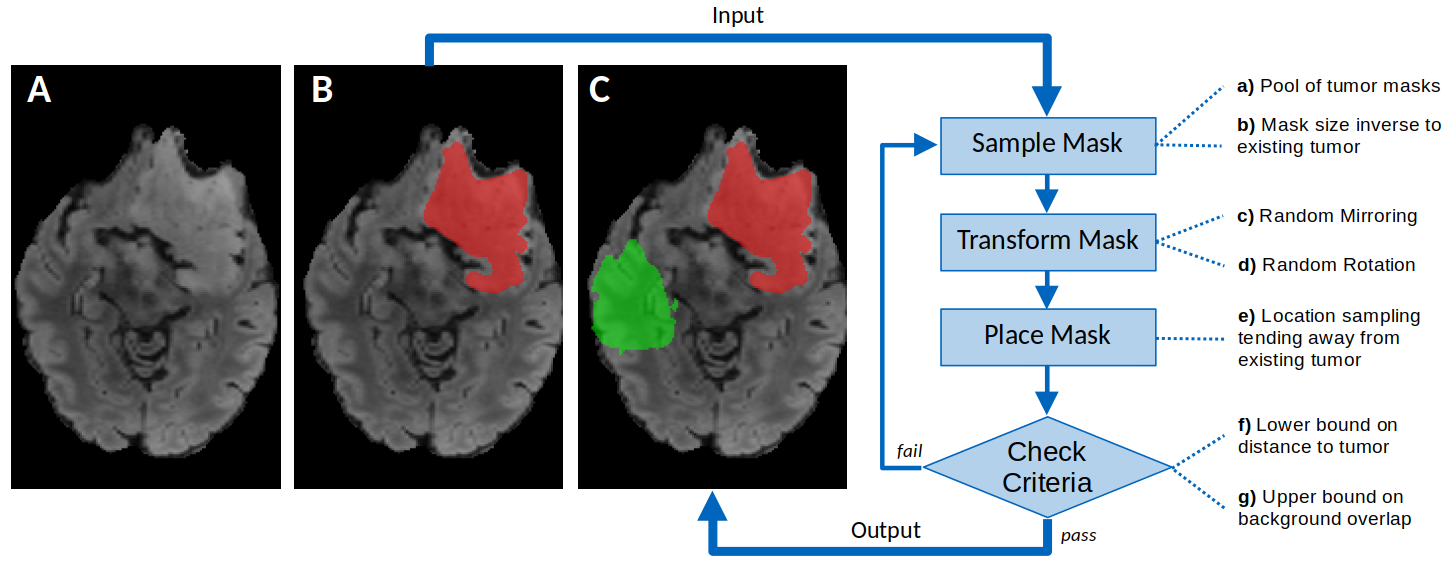}
    \caption{
        \textbf{Healthy inpainting masks in the RSNA-ASNR-MICCAI BraTS 2023 inpainting challenge:}
        Exemplary healthy mask generation procedure for one brain and one healthy mask.
        \textbf{Panel A} shows the underlying T1 scan and \textbf{panel B} has the respective tumor annotation overlaid in \textbf{red}).
        T1 scan and tumor annotation are the input for our mask sampling algorithm (\textbf{right side}.
        The algorithm randomly samples a mask, transforms it, and places it somewhere in the brain until a valid configuration is found.
        \textbf{Panel C} shows a valid example of a healthy mask (\textbf{green}) in the same brain as the tumor.
        These two different annotations are provided for the training set.
        The algorithms are supposed to inpaint both the masked tumor as well as the masked healthy area.
    }
    \label{fig:infill_annotations}
\end{figure}

%% file: body/definition/manual_curation.tex
\subsection{Manual Label Curation}
Besides the tumor tissue, patient brains frequently feature further pathologies.
As described in \Cref{sec:healthy_region_protocol} an algorithm proposes multiple candidates for healthy brain masks to evaluate the inpainting methods.
To ensure the areas proposed by the algorithm are indeed healthy, we conduct a manual expert assessment.
Therefore, for each case in our test set a trained medical expert determines the most suitable candidate mask.
Statistically, larger masks are more likely to contain pathologies, leading to a selection trend towards smaller healthy masks.
To counteract this trend, larger masks were preferred when possible.

\textbf{Glioma test set curation}:
The \href{https://www.synapse.org/Synapse:syn51156910/wiki/622351}{BraTS Glioma} test dataset comprises 570 patient exams.
In the first iteration, three healthy mask candidates were generated for all of the exams.
This allowed the medical expert to identify a healthy region for 546 patients.
In a second iteration, seven additional mask candidates were generated and reviewed for the remaining 24 patients.
This added 22 patients to the test set.
The remaining two patients were excluded due to pronounced cerebral tumor infiltration and multiple confluent lesions of other etiologies.
Overall, healthy masks were available for 568 patients.

\textbf{Meningioma test set curation}:
For the 2024 iteration of the inpainting challenge, an additional meningioma test dataset from the \href{https://www.synapse.org/Synapse:syn51156910/wiki/622353}{BraTS Meningioma Challenge} is introduced.

The dataset consists of 284 patient exams.
Similarly to the curation of the glioma dataset (see above), six candidate masks were generated for each exam.
Reducing the test set to 277 cases, the medical expert excluded seven patients due to pathologies; see Appendix \Cref{sec:AppendixManual}.

%% file: body/definition/participation.tex
\subsection{Participation}
The challenge commences with releasing the training dataset, consisting of imaging data, the corresponding inpainting masks, and the underlying tumor and healthy tissue masks.


Participants can start designing and training their methods using this training dataset.
The validation data is released within three weeks after the training data is released.
This allows participants to obtain preliminary results in unseen data and report these in their submitted short MICCAI LNCS papers, in addition to their cross-validated results on the validation data.

Finally, after uploading their containerized method to the evaluation platforms, all participants are evaluated and ranked on the same unseen testing data, which are inaccessible to the participants.
The final top-ranked participating teams are announced at the respective annual MICCAI Meeting.
The top-ranked participating teams in both tasks receive material prizes.

%% file: body/definition/baseline.tex
\subsection{Baseline Model}\label{sec:baseline}

Our baseline model implementation serves as a benchmark and illustration of the task at hand. \Cref{tab:2023results_baseline} depicts the performance of the baseline models on the glioma and meningioma test sets. The reference model can be accessed on the \href{https://github.com/BraTS-inpainting/2023_challenge/tree/main/baseline}{challenge GitHub site}, which utilizes a pix2pix architecture \citep{pix2pix}.
This model predicts solely the missing mask and concatenates it to the masked image.
pix2pix represents an autoencoder that utilizes an added \eac{GAN} loss to enhance image fidelity.
Performance evaluation for the baseline model will be done using the same measurements as mentioned in \Cref{sec:evaluation}.

\input{body/2023/results/tab/2023results_baseline}

%% file: body/2023/results/tab/2023results_baseline.tex
\begin{table}[H]
    \scriptsize
    \centering
    \begin{tabularx}{\textwidth}{ XXXX } 
        \toprule
        \textbf{Baseline Model} & \textbf{\textit{SSIM}} & \textbf{\textit{RMSE} } & \textbf{\textit{PSNR}}\\
        \midrule
         AutoEncoder & 0.85 [0.19] & 0.11 [0.04] & 19.39 [3.56] \\ 
         2Pix2Pix  & 0.85 [0.19] & 0.10 [0.05] & 19.68 [4.40] \\ 
        \bottomrule
    \end{tabularx}
    \caption{Evaluation of baseline model. All results of \eac{SSIM}, \eac{RMSE} and \eac{PSNR} are formatted as 'median [interquartile range]' \citep{sara2019image}}.
    \label{tab:2023results_baseline}
\end{table}

%% file: body/definition/evaluation/evaluation.tex
\subsection{Performance Evaluation} \label{sec:evaluation}
To measure the performance of the contributions, we evaluate the quality of the infilled regions.
Since \emph{ground truth} data is only available for the masked regions with healthy tissue, the evaluation will be restricted to the healthy masks.
We will use the following set of well-established metrics to quantify how realistic the synthesized image regions are compared to real ones: \eac{SSIM} \footnote{Usually, for \eac{SSIM}, the complete ground truth and inference image are compared against each other.
As we only want to evaluate inpainted regions, we take the voxel-wise \eac{SSIM} values and compute the mean over all voxels in the healthy mask.
}  \citep{wang2004image}, \eac{PSNR}, and \eac{RMSE}.
A \href{https://pypi.org/project/inpainting}{Python package} to compute these metrics is provided.

For the final ranking of the MICCAI challenge, an equally weighted rank-sum is computed across all three metrics.
To compute the rank for each metric, we apply a case-based ranking \citep{maier2018rankings, reinke2023eliminating} by ranking the participants for each test case and again compute a rank-sum, as illustrated in \Cref{fig:ranking_scheme}. As proposed by \citep{wiesenfarth2021methods}, we analyzed ranking robustness by bootstrapping the ranking across 1,000 bootstrap samples to analyze how stable the rankings were against small perturbations within the test dataset. Furthermore, we calculated pairwise significance tests using the one-sided Wilcoxon signed rank test at a 5\% significance level with adjustment for multiple testing according to Holm. 
The rankings and their robustness analysis are computed \href{https://github.com/BraTS-inpainting/2023_challenge_ranking/tree/main/ranking_scheme}{using challengeR} \citep{wiesenfarth2021methods}.

\input{body/definition/evaluation/fig/ranking_scheme}

%% file: body/definition/evaluation/fig/ranking_scheme.tex
\begin{figure}[htbp!]
        \centering
        \includegraphics[width=0.8\linewidth]{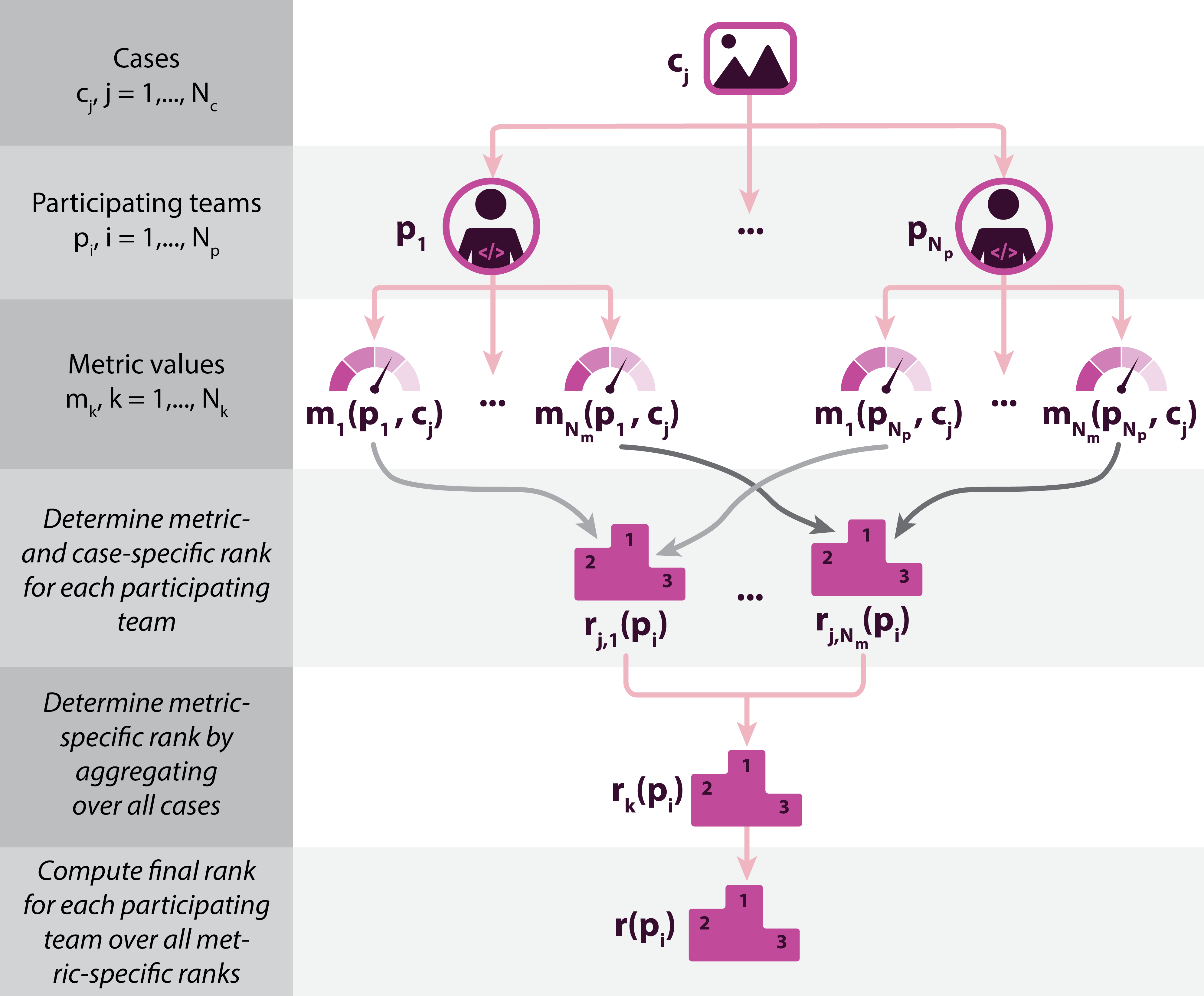}
        \caption{
            \textbf{Ranking Scheme of the challenge.}
                First, a case-based ranking is calculated for every metric, i.e., for each test case $c_j, j=1,\cdots,N_c$, the performance $m_k(p_i,c_j)$ is calculated for each participating team $p_i, i=1,...,N_p$ and for each metric $m_k\in \{\text{\eac{RMSE}}, \text{\eac{PSNR}}, \text{\eac{SSIM}}\}$. Based on $m_k(p_i,c_j)$, a metric- and test-case specific rank $r_{j,k}(p_i)$ is calculated for each participating team. If $m_k(p_i,c_j) = \texttt{NA}$, $r_{j,k}(p_i)$ is set to the last rank. A metric-specific rank $r_k(p_i)$ is determined by aggregating the ranks over all cases. Second, the final rank for each participating team $r(p_i)$ is calculated over all metric-specific ranks. 
        }
        \label{fig:ranking_scheme}
\end{figure}


%% file: body/2023/2023.tex
\section{2023 challenge}
\label{sec:2023challenge}

\input{body/2023/submissions/submissions}

\input{body/2023/results/results}

\input{body/2023/ranking/ranking_robustness}

%% file: body/2023/submissions/submissions.tex
More than 250 teams signed up for the 2023 BraTS inpainting challenge.
Out of these, nine submitted a manuscript describing their inpainting method; however, two papers were withdrawn.
Of the remaining seven teams, five submitted a containerized version of their algorithm as an \href{https://mlcommons.org/working-groups/data/mlcube/}{MLCube}; one team withdrew their submission.
Thanks to the sponsors, we could award 500 USD to first place, 250 USD to second place, and 125 USD to third place.
The four remaining teams competed for this prize money.
The participants selected very different approaches to the inpainting task, as summarized by \Cref{tab:2023submissions}.

\input{body/2023/submissions/tab/2023submissions}

%% file: body/2023/submissions/tab/2023submissions.tex
\begin{table*}[h!]
    \scriptsize
    \centering
    \begin{tabularx}{\textwidth}{X p{0.15\textwidth} XXX} 

        \toprule
        \textbf{Authors} & \textbf{Dimensionality} & \textbf{Model type} & \textbf{Loss function } & \textbf{Inference speed}\\
        \midrule
         VisionLabODU23 et al. & 3D  & Pix2Pix & perceptual, adversarial & Fast \\
         MedSegCTRL et al. & 2D  & Big-LaMa & \eac{MAE}, perceptual, adversarial, feature matching & Intermediate \\
         Domaso et al. & 2D  & Diffusion model & \eac{MSE} & Slow \\
         Ying-Weng et al. & 3D  & U-Net & \eac{MAE}, \eac{SSIM} & Fast \\
        \bottomrule
    \end{tabularx}
    \caption{Comparison of approaches employed in the 2023 inpainting challenge.
    The table displays dimensionality, model type, loss function, and interference speed of each model.
    Challenge participants employed a wide variety of approaches, showing that the community did not converge to a single approach to inpainting yet.
}
    \label{tab:2023submissions}
\end{table*}

%% file: body/2023/results/results.tex
\subsection{Results}
\Cref{tab:2023metrics} summarizes the quantitative scores of the submissions, while \Cref{fig:violin_plots} shows the distribution of metric scores for all teams as violin plots.
Across all quantitative metrics, the method by Ying Weng et al. performed best. 
Nevertheless, qualitatively inspecting the generated images painted a more faceted picture, c.f. \Cref{fig:qualitativ_comparison}.

\input{body/2023/results/tab/2023results}

\input{body/definition/evaluation/fig/violin}

\input{body/2023/results/fig/qualitativ_comparison}

%% file: body/2023/results/tab/2023results.tex
\begin{table*}[h]
    \scriptsize
    \centering
    \begin{tabularx}{\textwidth}{ p{2 cm} p{4.1 cm} p{3.1 cm} p{3.1 cm} p{3.1 cm} }
        \toprule
         \textbf{Ranking} & \textbf{Participants} & \textbf{ \eac{SSIM} } & \textbf{ \textit{RMSE} } & \textbf{ \eac{PSNR} } \\  
        \midrule
         1 & Ying-Weng et al. & 0.91 [0.15] & 0.07 [0.04] & 23.59 [5.35] \\
         2 & Domaso et al. & 0.86 [0.20] & 0.10 [0.04] & 20.42 [3.82] \\
         3  & MedSegCTRL et al. & 0.87 [0.18] & 0.12 [0.05] & 18.71 [4.01] \\
         4  & \textit{withdrawn} & 0.79 [0.31] & 0.12 [0.16] & 18.33 [9.75] \\
         5  & VisionLabODU23 et al. & 0.77 [0.24] & 0.15 [0.06] & 16.30 [3.53] \\
        \bottomrule
    \end{tabularx}
    \caption{
    Evaluation metrics of the top-5 contributions to the 2023 inpainting challenge.
    We report median and interquartile range for each evaluation metric.
    }
    \vspace{-0.5cm}
    
    \label{tab:2023metrics}
\end{table*}


%% file: body/definition/evaluation/fig/violin.tex
\begin{figure}[htbp!] 
        \centering
        \includegraphics[width=1.0\linewidth]{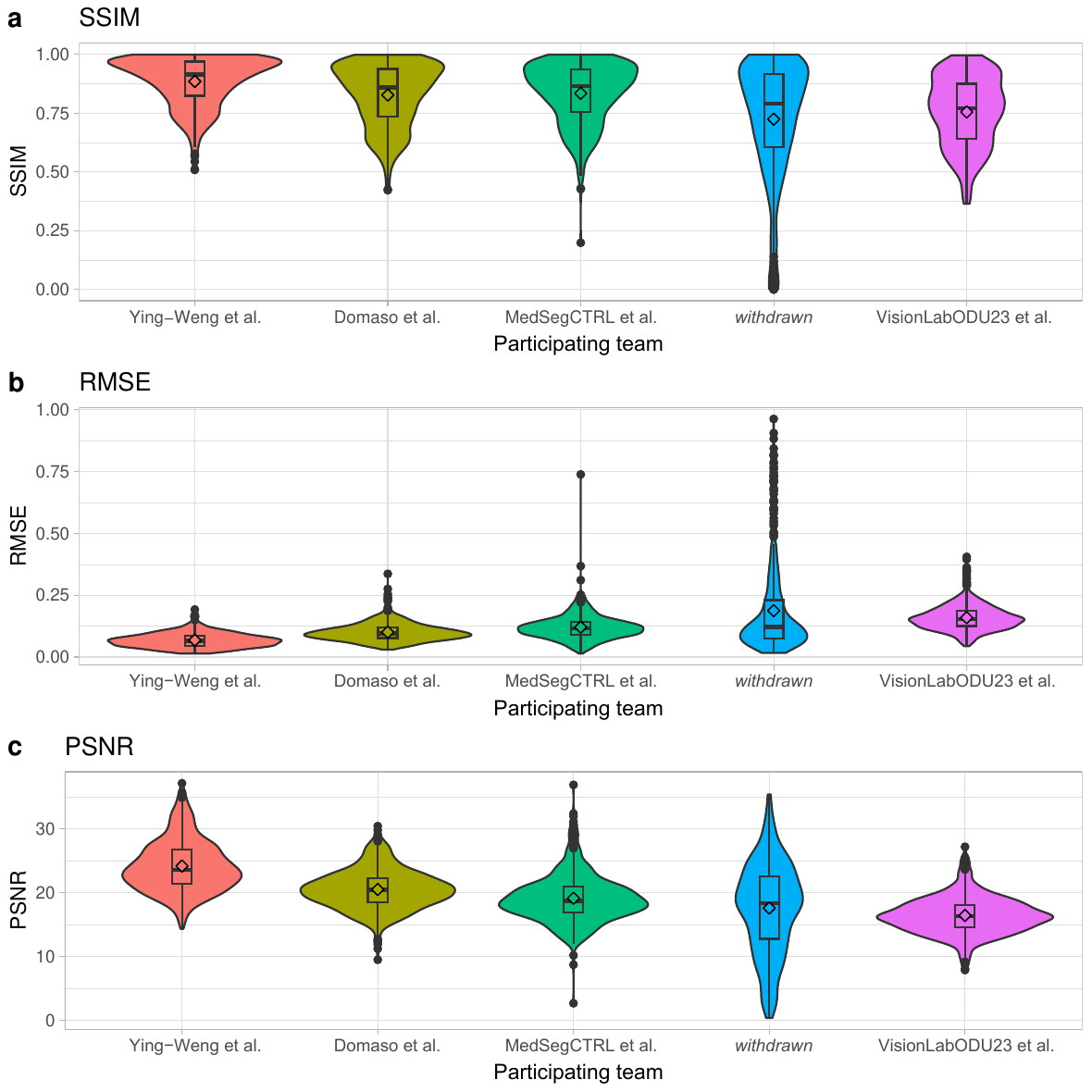}
        \caption{
            \textbf{Violin plots showing the individual performance of participating teams.}
                Results are shown for (a) \eac{SSIM}, (b) \eac{RMSE}, and (c) \eac{PSNR}. Median performance is indicated by a thick horizontal line, mean performance by a rhombus. The upper and lower border of the boxplots illustrate the first and third quartile. The density of individual metric scores is shown by the violin plot.
        }
        \label{fig:violin_plots}
\end{figure}

%% file: body/2023/results/fig/qualitativ_comparison.tex
\begin{figure}[htbp]
        \centering
        \includegraphics[width=1.0\linewidth]{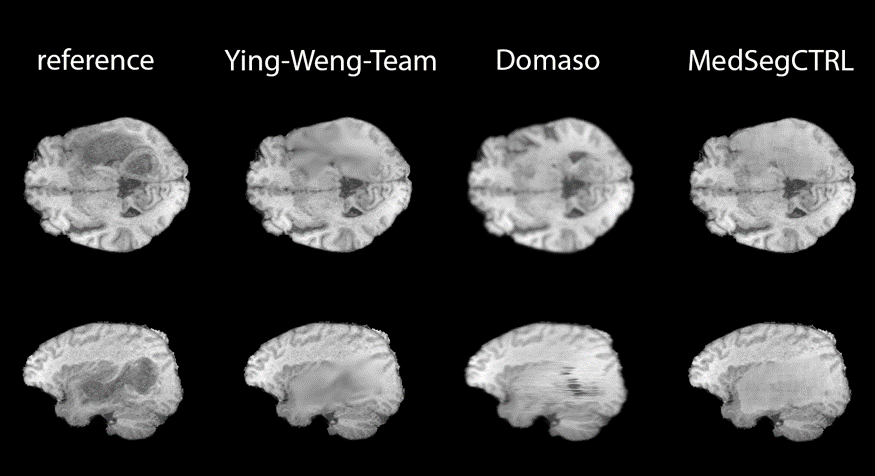}
        \caption{
            \textbf{Qualitative comparison of inpainting sections from the 2023 challenge.}
            Reference section (left) and the top three ranking teams with their respective in-painted solutions in axial (upper row) and sagittal plane (lower row).
            The 2D diffusion model from team Domaso produces visually-pleasing images on the axial slices.
            However, stripe artifacts appear in the sagittal and coronal (not displayed) view.
            In contrast the other methods produce blurry inpaintings across all three views.
        }
        \label{fig:qualitativ_comparison}
\end{figure}

%% file: body/2023/ranking/ranking_robustness.tex
\subsection{Ranking robustness analysis}
From the violin plots in \Cref{fig:violin_plots}, it is already visible that the rankings seems quite clear. This is further confirmed by applying bootstrapping and analyzing the ranking robustness directly, as shown in \Cref{fig:blob_plots}. The figure illustrates how often every participating team achieved ranks 1-5 across the 1,000 bootstrap rankings. It can be seen that the ranking is extremely stable. Teams Ying-Weng et al. (rank 1), Domaso et al., (rank 2) and VisionLabODU23 et al. (rank 5) receive their original ranks across all bootstrap samples. Only team MedSegCTRL et al. and the withdrawn submission exchange their ranks in some rare occurences while sticking to their original ranks in most of the times. This is confirmed by computing the rank correlation coefficient Kendall's tau across bootstrap samples, which has a mean, median, and 25\% and 75\% percentiles of 1.0, indicating an extremely stable ranking.

Similar results are also perceived when comparing how often one team is significantly superior in performance to the others (see \Cref{app:sign_maps}, \Cref{fig:significance_maps}). In nearly all cases, the teams are significantly superior to all their lower-ranked teams.

\input{body/definition/evaluation/fig/blob}

%% file: body/definition/evaluation/fig/blob.tex
\begin{figure}[ht!]
        \centering
        \includegraphics[width=1\linewidth]{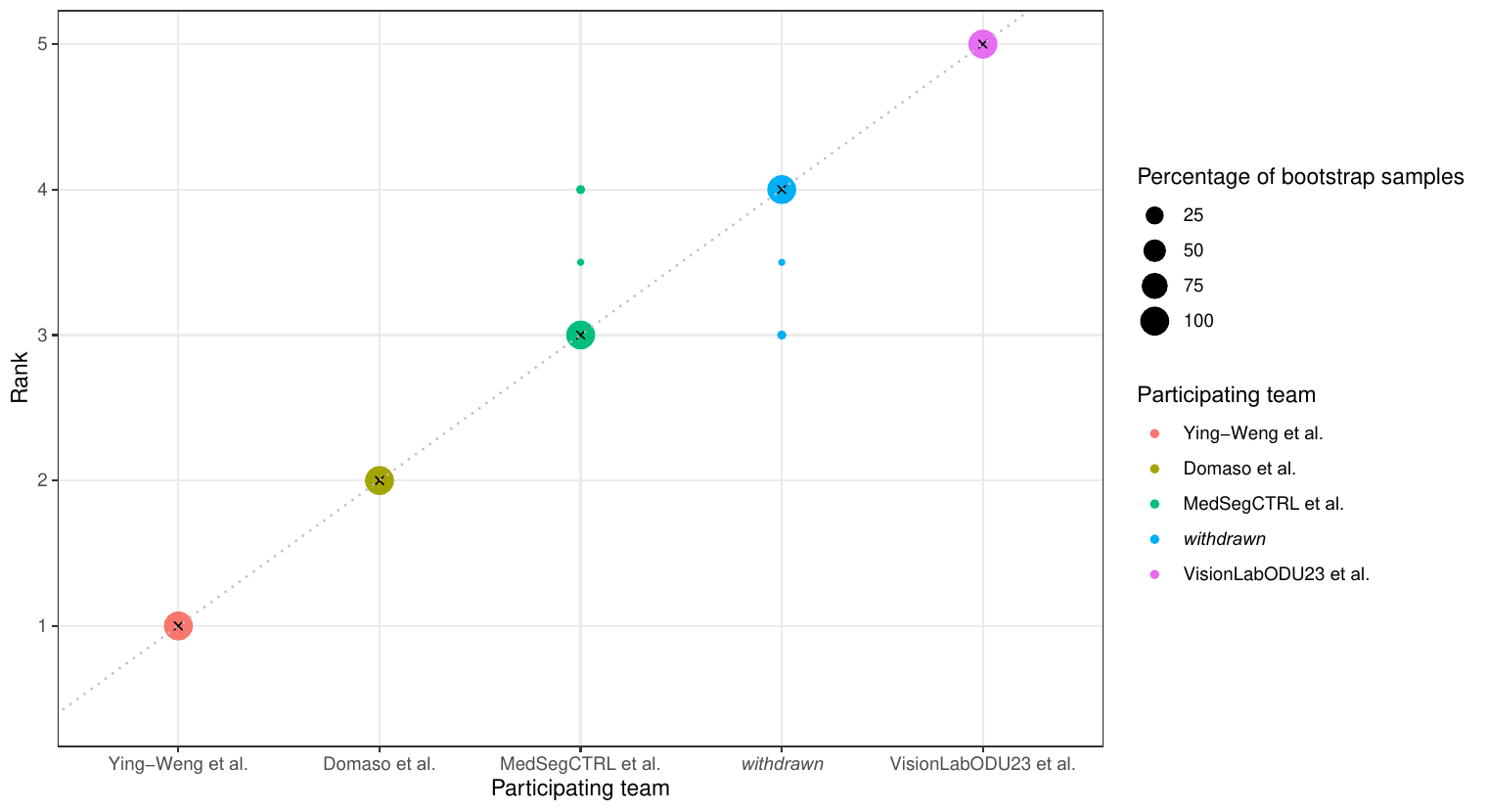}
        \caption{
            \textbf{Blob plots showing that the challenge ranking is robust against small perturbations.}
                The radius of the blobs, which are color-coded by the participating teams, indicates the frequency of each team achieving a specific rank across 1,000 bootstrap samples. The median rank of each participating team is indicated by a cross. The black lines show the 95\% bootstrap intervals across the 1,000 bootstrap samples.
        }
        \label{fig:blob_plots}
\end{figure}

%% file: body/2024/2024.tex
\section{2024 challenge}
\label{sec:2024challenge}
The inpainting challenge is repeated for MICCAI 2024.
The overall challenge design, more specifically, task, evaluation metrics, and ranking scheme, remain the same.
In addition to the glioma test set, a second test set with 277 patients from the BraTS meningioma challenge is introduced to investigate how well the algorithms generalize toward other pathologies and image acquisition with different \eac{MRI} scanners.
More information is available on the \href{https://www.synapse.org/Synapse:syn53708249/wiki/627498}{challenge website}.

%% file: body/discussion/discussion.tex
\section{Discussion}


\input{body/discussion/2023_learnings}

%% file: body/discussion/2023_learnings.tex
A few learnings can be derived from the 2023 inpainting challenge.
First, inpainting represents a small but important niche in the biomedical image analysis community.
There are significant barriers to participating in the challenge, as training inpainting models quickly becomes computationally and, therefore, financially expensive.
The strategy of directly optimizing evaluation metrics as outlined by \citet{maier2018rankings} also represents a fruitful approach here.
Previous research demonstrated that human expert perception is not fully aligned with established \eac{ML} metrics for segmentation tasks \citep{melba:2023:002:kofler}.
This seems at least equally true for image synthesis tasks; consequently, there is an apparent need for innovative image synthesis metrics that better align with human (expert) perception to evaluate paintings \citep{kofler2023approaching}.
Therefore, human surrogate models \citep{kofler2022deep} might complement the existing metrics in future iterations of the challenge to achieve a more comprehensive evaluation.
Last, judging from the qualitatively-observed outperformance of 2D diffusion models on axial slices, c.f. \Cref{fig:qualitativ_comparison}, 3D diffusion models seem promising to achieve human-pleasing inpaintings.



%% file: body/acknowledgement/acknowledgement.tex
\section{Acknowledgement}
S.B. was supported by the National Institutes of Health (\eac{NIH}) under the award number NIH/NCI:U01CA242871. The content of this publication is solely the responsibility of the authors and does not represent the official views of the \eac{NIH}.

%% file: appendix/appendix.tex
\section{Ranking stability based on statistical significance}
\label{app:sign_maps}
\input{appendix/fig/significance_maps}

\section{Manual Curation}\label{sec:AppendixManual}
\input{appendix/manual_curation}

%% file: appendix/fig/significance_maps.tex
\begin{figure}[ht!]
        \centering
        \includegraphics[width=1\linewidth]{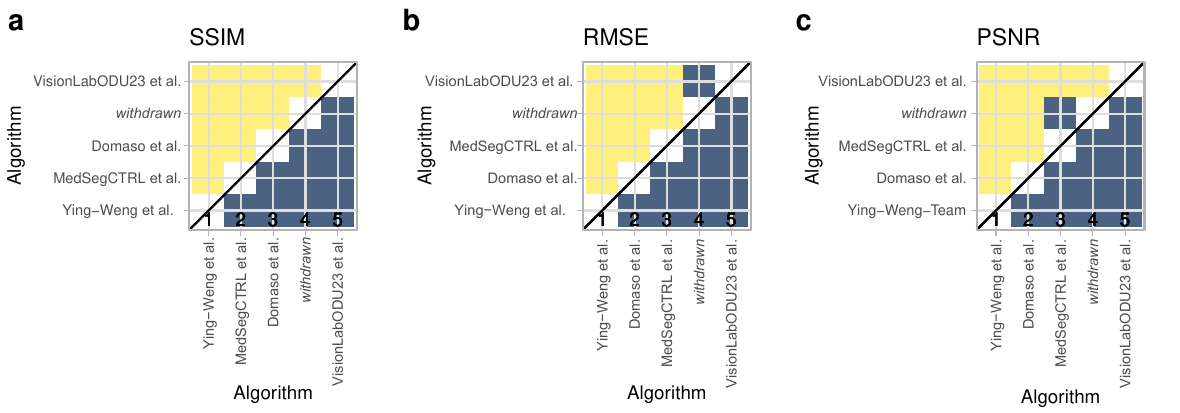}
        \caption{
            \textbf{Significance maps for visualizing ranking robustness based on statistical tests.}
                Results are shown for (a) \eac{SSIM}, (b) \eac{RMSE}, and (c) \eac{PSNR}. The plots show the incidence matrices of pairwise significant test outcomes. Yellow squares indicate that metric values from the participating team on the x-axis were significantly superior to those on the y-axis, while blue shading denotes no significant difference. 
        }
        \label{fig:significance_maps}
\end{figure}

%% file: appendix/manual_curation.tex
The higher exclusion rate in the Meningioma dataset may be due to a potentially different distribution of tumor sizes between the two datasets. 
The size distribution of Meningioma in our datasets seems to have a higher variance than the Glioma dataset (i.e. more small tumors and more big tumors). Our mask generation algorithm tends to choose bigger healthy masks when the actual tumor is small (and vice versa, see chapter \Cref{sec:healthy_region_protocol}. As a result, for a higher variation in actual tumor sizes (Meningioma) the size ratio between the actual tumor and the generated healthy mask is also more variable (i.e. more huge tumors with tiny healthy candidates and more tiny tumors with huge healthy candidates). Because large healthy candidate masks are more likely to also cover other cerebral pathologies, the likelihood of a candidate being indeed healthy is smaller. Consequently, the increased variance of the Meningioma size distribution results in fewer valid healthy mask candidates.